\documentclass[a4paper,aps,preprintnumbers,showpacs,twocolumn,superscriptaddress,nofootinbib]{revtex4-1}

\usepackage{amsmath}
\usepackage{amssymb}
\usepackage{graphicx}
\usepackage{epsfig}
\usepackage{color}
\usepackage{url}
\usepackage{times}
\usepackage{bm}
\usepackage{mathrsfs}
\usepackage[utf8]{inputenc}
\usepackage{hyperref}
\usepackage{enumerate}
\usepackage{amsthm}
\usepackage{bbm}

\newcommand{\beq}{\begin{equation}}
\newcommand{\eeq}{\end{equation}}
\newcommand{\bea}{\begin{eqnarray}}
\newcommand{\eea}{\end{eqnarray}}
\newcommand{\bit}{\begin{itemize}}
\newcommand{\eit}{\end{itemize}}
\newcommand{\ben}{\begin{enumerate}}
\newcommand{\een}{\end{enumerate}}
\newcommand{\nn}{\nonumber}

\def\scri{\mathscr{I}}

\newcommand{\QMUL}{\affiliation{School of Mathematical Sciences, Queen Mary, University of
  London, \\ Mile End Road, London E1 4NS, United Kingdom}}
          
\theoremstyle{plain}
\newtheorem{proposition}{Proposition}

\newtheorem{remark}{Remark}

\begin{document}
\title{Comment on ``Some exact quasinormal frequencies of a massless scalar field in Schwarzschild spacetime"
}
\author{Rodrigo Panosso Macedo} 
\QMUL
\date{\today}

\begin{abstract}
A new branch of quasinormal modes for a massless scalar field propagating on the Schwarzschild spacetime was recently announced~\cite{Batic2018}. We review the quasinormal modes characterisation and arguments and identify the flaws in their proof. Then, we preset explicit counterexamples to such arguments. Finally, we study the modes via alternative methods and do not find the new branch. We conclude against their interpretation.
 \end{abstract}

\maketitle
{\it Introduction.}
Recently,Ref.~\cite{Batic2018} announced a new branch of quasinormal modes (QNMs) for the scalar field on the Schwarzschild spacetime (see~\cite{Chandrasekhar83,Kokkotas99a,Nollert99,Berti:2009kk,Konoplya:2011qq} for main reviews). That study is based on Leaver's work~\cite{Leaver85}. The approach deals with a $3$-term recurrence relation and the claimed QNMs correspond to values $s^{\rm new}_k\neq s^{\rm QNM}_k$ where one coefficient of the recurrence relation vanishes. In~\cite{Ansorg2016,Macedo2018}, Leaver's approach is reinterpreted in the context of a hyperboloidal foliation of the spacetime and the algorithm is extended to discuss the spectral decomposition of the solution to the wave equation. 

Here, we present the flaws in arguments in Ref.~\cite{Batic2018}, which are then identified in counterexamples. Specifically, the forward iteration of the recurrence relation leads to an exponential grow of the coefficients. While this growth agrees with the theoretical prediction, it contradicts a central assumption in the authors' arguments that the sequence should decay. Complementarily, we calculate the QNMs as the eigenvalues of the operator associated to the wave equations and its boundary conditions and study its explicit time evolution. In neither of the alternatives approaches do we find the new frequencies.

\medskip
{\it Scalar field on Schwarzschild.}
The wave equation for a massless scalar field propagating on the Schwarzschild background reads (units $c=G=1$)
\bea
\label{eq:WaveEq}
&-f_{,\bar{t}\bar{t}} + f_{,x x} - {\cal P}f = 0,\\ 
\label{eq:Potential}
&{\cal P} = \left( \dfrac{2M}{r} \right)^{2}\left(1-\dfrac{2M}{r}\right)\left[\ell(\ell+1) + \dfrac{2M}{r} \right].
\eea
Here, $\bar{t}=t/(2M)$ and $x= r/(2M) + \ln\left[r/(2M) - 1\right]$ are, respectively, the dimensionless time and tortoise coordinate.

We introduce~\cite{Ansorg2016} the hyperboloidal coordinates $\{\tau, \sigma, \theta, \varphi\}$
\beq
\label{eq:HypCoords}
t = 2M\bigg( 2\tau +\sigma^{-1} - \ln\left[ \sigma(1-\sigma)\right]\bigg), \quad r = 2M/\sigma.
\eeq
The black-hole horizon ${\cal H}^+$ is at $\sigma = 1$, whereas $\sigma = 0$ locates future null infinity $\scri^+$.
The equation for $V(\tau,\sigma) = f(t(\tau,\sigma), r(\sigma))$ reads
\bea
\label{eq:WaveEqHyp}
&-(1+\sigma)V_{,\tau \tau} + (1-2\sigma^2) V_{,\tau\sigma} + \sigma^2(1-\sigma)V_{,\sigma\sigma} \nn \\
&+\sigma[2-3\sigma]V_{,\sigma} - 2\sigma V_{,\tau} - [\ell(\ell+1) + \sigma] =0.
\eea
The regularity conditions at $\scri^+$ and ${\cal H}^+$ account for the desired physical boundary conditions~\cite{Ansorg2016,Macedo2018}.

The Laplace transform of Eq.~\eqref{eq:WaveEqHyp} yields $ {\mathbf A}(s) \hat{V}(s) = B(s)$. The source term $B(s)$ contains the initial data~\cite{Ansorg2016,Macedo2018}, and \bea
&{\mathbf A}(s)=\sigma^{2}(1-\sigma)\partial_{\sigma\sigma} +\left\{ s(1-2\sigma^2)+\sigma\left[2-3\sigma \right]\right\}\partial_{\sigma} \nn \\
& -\left[s^{2}(1+\sigma) + 2\sigma\,s  + \ell(\ell+1) + \sigma\right].
\eea
For the QNMs, we focus on the homogenous equation
\beq
\label{eq:HomEq}
{\mathbf A}(s) \phi(s) = 0.
\eeq
This particular hyperboloidal foliation $\{\tau, \sigma\}$ is the spacetime counterpart of the analysis in Ref.~\cite{Leaver85}. Indeed, substituting
\beq
\label{eq:TimeDepAnsatz}
f(\bar{t}, x) =  e^{\bar{s}\bar{t}} F(x;\bar{s}) 
\eeq
into~\eqref{eq:WaveEq}, we obtain $F_{,xx} - \left[ \bar{s}^2 + {\cal P}\right]F = 0,$
with $s = 2\bar{s}.$
Ingoing/outgoing boundary conditions at infinity impose
\bea
\label{eq:LaplCauchyBoundary}
F (x;\bar{s}) &\sim& e^{\mp\bar{s}x}, \quad x\rightarrow \pm \infty.
\eea
While Leaver incorporates such asymptotics via~\cite{Leaver85}
\beq
\label{eq:Rel_f_V}
F(x(\sigma);\bar{s}) = \Xi(\sigma;\bar{s}) \,\phi(\sigma; 2\bar{s}), \,\,
\Xi(\sigma; \bar{s}) = 2 e^{-\bar{s}/\sigma}\sigma^{\bar{s}}(1-\sigma)^{\bar{s}},
\eeq
in Refs.~\cite{Ansorg2016,Macedo2018}, $\Xi(\sigma; \bar{s})$ follows directly from the substitution of~\eqref{eq:HypCoords} into~\eqref{eq:TimeDepAnsatz}. 
Finally, we expand $\phi(\sigma;s)$ in the Taylor series
\beq
\label{eq:HomSol_H}
\phi(\sigma; s) = \sum_{k=0}^{\infty} H_k (1-\sigma)^k.
\eeq
Eq.~\eqref{eq:HomSol_H} into \eqref{eq:HomEq} gives the $3$-term recurrence relation 
\beq
\alpha_k H_{k+1} + \beta_{k} H_{k} + \gamma_{k} H_{k-1} = 0 \quad (k\ge 1), \label{eq:RecRel}
\eeq
with $\alpha_k = (k+1)( k+1 + s ),$ $-\beta_k = 2(k+s)( k+1 + s) + \ell(\ell+1)  + 1$ and $\gamma_k = (k+s)^2$. 

\medskip
{\it Initial conditions.} The initial seeds $H_0$ and $H_1$ for \eqref{eq:RecRel} must satisfy the initial condition
\beq
\alpha_0 H_{1} + \beta_{0} H_{0}  = 0 \label{eq:RecRel_BC}. \\
\eeq
Equation~\eqref{eq:RecRel_BC} ensures the regularity of $\phi(\sigma; s)$ at $\sigma=1$, and it is equivalent to \eqref{eq:RecRel} at $k=0$ if and only if $H_{-1} =0$. 

\begin{proposition}
\label{rem:ForwardIteration}
For {\em any} fixed $s\neq {\mathbb Z}^-$, there exists a unique (up to normalisation) sequence $\{H_k\}_{k=0}^\infty$ satisfying both \eqref{eq:RecRel_BC} and \eqref{eq:RecRel}. A convenient normalisation is $H_0=1$.
\end{proposition}

To construct $\{H_k\}_{k=0}^\infty$, one starts with $H_0=1$, then obtains $H_1=-\beta_0/\alpha_0$ according to~\eqref{eq:RecRel_BC}, and finally calculates the remaining $H_k$ ($k\ge1$) via a forward iteration of \eqref{eq:RecRel}. 

For a given {\em exact} value $s$, the forward iteration of Eqs.~\eqref{eq:RecRel_BC} and \eqref{eq:RecRel} is an exact calculation without any source numerical error. The technical limitation is restricted to a truncation $k=N_{\rm max}$ in the forward iteration procedure. From the practical perspective, one always deal with a finite sequence $\{H_k\}_{k=0}^{N_{\rm max}}$. For large $N_{\rm max}$, the values must be confronted against the asymptotic behaviours of the solutions to~\eqref{eq:RecRel}.

\medskip
{\it Asymptotic behavior.} Equation~\eqref{eq:RecRel} admits two independent asymptotic behaviours~\cite{Batic2018,Ansorg2016}
 \bea
 \label{eq:ExpGrowDec}
&  H_{k,\pm} \sim k^\zeta  \, e^{\pm\kappa\sqrt{k}}A_{k,\pm}, \quad \Re(\kappa)>0, \\
 & \kappa=2\sqrt{s}, \quad \zeta= {s}/{2}-{3}/{4},   \quad A_{k,\pm} = 1 + \sum\limits_{j=1}^{\infty}\dfrac{\mu_{\pm,i}}{k^{j/2}}. \nn
 \eea
\begin{proposition}
\label{rem:BackIteration}
For {\em any} fixed $s\neq {\mathbb Z}^-$ there exists unique (up to normalisation) sequences $\{H_{k,+} \}_{k=0}^\infty$ and $\{H_{k,-} \}_{k=0}^\infty$ satisfying \eqref{eq:RecRel} and the asymptotic behaviour~\eqref{eq:ExpGrowDec}. A convenient normalisation is $\displaystyle \lim_{k\rightarrow \infty} k^{-\zeta} e^{\mp\kappa\sqrt{k}} H_{k,\pm} = 1$.
\end{proposition}
Thus, $\{H_k\}_{k=0}^\infty$ from Proposition~\eqref{rem:ForwardIteration} is a linear combination,
\beq
\label{eq:Hk_Asymptotics} 
H_k =  \lambda_+\, H_{k,+} + \lambda_-\, H_{k,-} .
 \eeq
 \begin{remark}
\label{rem:ForwardIteration_Asympt}
By imposing the initial seeds~\eqref{eq:RecRel_BC} at a given values $s$, a unique solution $\{H_k\}_{k=0}^\infty$ to ~\eqref{eq:RecRel} is fixed. Thus, one does not have the freedom to choose a given asymptotic behaviour in the linear combination~\eqref{eq:Hk_Asymptotics}. In general, the exponential growth $H_k \sim  e^{+\kappa\sqrt{k}} k^\zeta$ dominates for large $k$.
\end{remark}

We are interested in the decaying solution $\{H_{k,-} \}_{k=0}^\infty$. While Proposition \ref{rem:BackIteration} is a formal result arising from an asymptotic study of Eq.~\eqref{eq:RecRel}, Refs.~\cite{Ansorg2016,Macedo2018} discuss its explicit calculation (closely related to Leaver's continued fraction~\cite{Leaver85}). Briefly, one truncates the series at a given value $k=N_{\rm max}$ and approximates the values $H_{N_{\rm max}, -}$ and $H_{N_{\rm max}-1, -}$ according to the behaviour \eqref{eq:ExpGrowDec}. The complete sequence $\{ H_{k,-} \}_{k=0}^{N_{\rm max}}$ is obtained by iterating~\eqref{eq:RecRel_BC} backwards up until $k=0$~\cite{Ansorg2016}. The backward iteration provides us with $H_{-1,-}$ as well, which plays a crucial role when asserting the validity~\eqref{eq:RecRel_BC}.

\begin{remark}
\label{rem:BackIteration_InitialCondition}
By imposing the decaying asymptotic behaviour in~\eqref{eq:ExpGrowDec} at a fixed values $s$, a unique solution $\{H_{k,-}\}_{k=0}^\infty$ to~\eqref{eq:RecRel} is fixed.
Thus, one does not have the freedom to impose the initial condition~\eqref{eq:RecRel_BC}. In general, one obtains $H_{-1,-}\neq 0$.
\end{remark}

At the QNMs $s^{\rm QNM}_n$~\cite{Leaver85} the sequence $\{H_k\}_{k=0}^\infty$ does decay asymptotically as approximately $k^\zeta  \, e^{-\kappa\sqrt{k}}$. The growing behaviour is absent in~\eqref{eq:Hk_Asymptotics} and one has $H_k = \lambda_- \, H_{k,-}$. If one starts with the decaying solution $\{ H_{k,-} \}_{k=0}^\infty$, the values $H_{0,-}$ and $H_{1,-}$ satisfy the initial condition~\eqref{eq:RecRel_BC}, i.e.,~$H_{-1,-} = 0$.

We finish this section with a final remark that fixes the notation used from now on:
 \begin{remark}
 \label{rem:Notationpm}
 Let a pair of complex conjugates $s$ values with $\Re(s)\le0$ be parametrized by
 \beq
\label{eq:QNMCandidate}
s^{(\pm)} = \rho^2 \, e^{\pm 2\phi\, i }, \quad \rho > 0, \quad \phi\in[\pi/4,\pi/2).
\eeq
Let $\{H^{(\pm)}_k\}_{k=0}^\infty$ be the respective sequences arising from Proposition \ref{rem:ForwardIteration}. Then, $H^{(+)}_k = \overline{H^{(-)}_k }$ and therefore
 \beq
 \label{eq:Asymt_y}
y_{k,\pm} := \left| \frac{H^{(+)}_{k+1}}{H^{(+)}_k}\right|=\left| \frac{H^{(-)}_{k+1}}{H^{(-)}_k}\right| = 1 \pm \frac{2\rho\cos\phi}{\sqrt{k}} + {\cal O}(k^{-1}).
\eeq
 \end{remark}

Here, $\bullet^{(\pm)}$ refers to quantities constructed out of the pair of complex conjugate values $s^{(\pm)}$,
whereas $\bullet_{\pm}$ is related to the two possible asymptotic in~\eqref{eq:ExpGrowDec} and \eqref{eq:Asymt_y}. Remark 3 emphasises that the two asymptotic behaviours $y_{k,\pm}$ are by no means
controlled by the sign of $\Im(s^{(\pm)})$. The authors in Ref.~\cite{Batic2018} acknowledge this
property in a sentence after the Eq. (29) in Ref.~~\cite{Batic2018}: "For both cases there
are always one exponentially increasing solution and one exponentially decreasing"~~\cite{Batic2018}.

\medskip
{\it Quasinormal modes.} For candidates, Ref.~\cite{Batic2018} considers
\beq
s^{(\pm)}_n = -n -\frac{1}{2} \pm \frac{i}{2}\sqrt{1+2\ell(\ell+1)}, \quad n\ge0. \label{eq:NewQNM}
\eeq 
According to Ref.~\cite{Batic2018}, the following conditions are met at a QNM:
\ben[I.]
\item The boundary conditions \eqref{eq:LaplCauchyBoundary} are fulfilled.
\item The recurrence relation \eqref{eq:RecRel} has a minimal solution. If so, convergence can be checked by the Gauss criterion.
\item The initial condition \eqref{eq:RecRel_BC} is satisfied and the recurrence relation \eqref{eq:RecRel} should not give rise to an under-/overdetermined system for the coefficients.
\een

Condition (I) imposes the appropriate boundary conditions leading to QNMs. By construction, they are taken into account via Eq.~\eqref{eq:HypCoords} (time domain) or Eq.~\eqref{eq:Rel_f_V} (frequency domain). We recall that Eqs.~\eqref{eq:LaplCauchyBoundary} are necessary, but not sufficient conditions for the existence of a QNM~\cite{Noller92,Nollert99,Kokkotas99a}. \cite{Ansorg2016} shows that (I) is satisfied for any $s$ with $\Re(s)<0$. Proposition ~\ref{rem:BackIteration} ensures the existence of the decaying sequence $\{H^{(\pm)}_{k,-} \}_{k=0}^\infty$ which addresses condition (II). Proposition~\ref{rem:ForwardIteration} ensures the existence of the sequence $\{H^{(\pm)}_{k} \}_{k=0}^\infty$ which fulfills (III). At a QNM one must verify that $\{H^{(\pm)}_{k} \}_{k=0}^\infty$ and $\{H^{(\pm)}_{k,-} \}_{k=0}^\infty$ are linearly dependent, i.e., that (II) and (III) are met for the same sequence.

Batic et at.~\cite{Batic2018} classify the values $s^{(-)}_n$ in Eq.~\eqref{eq:NewQNM} as QNMs. Reference~\cite{Batic2018} discusses condition (II) by rewriting the (generic) asymptotic behaviour~\eqref{eq:Asymt_y} in terms of the (specific) values~\eqref{eq:NewQNM}. There is a clear {\em choice} for the decaying behaviour $y_{k,-}$: ``the case with the plus sign in" \eqref{eq:Asymt_y} ``can be disregarded [...]. Hence, the only relevant case to be considered is the one for" $H^{(\pm)}_{k,-}$~\cite{Batic2018}. This choice is equivalent to using Proposition~\ref{rem:BackIteration}, which ensures the existence of the unique asymptotically decaying solutions $\{H^{(\pm)}_{k,-}\}_{k=0}^\infty$ at the values $s^{(\pm)}_n$ --- cf. Eq.~\eqref{eq:NewQNM}.

In Ref.~\cite{Batic2018}, the choice for the decaying solution is attached to a restriction to the values
$s_n^{(-)}$, i.e., for the authors, the decaying behaviour of the sequence $\{ H^{(-)} _{\pm}\}$
is a direct consequence of the negative imaginary part of $s_n^{(-)}$. This line of reasoning is wrong, and it contradicts their own arguments in Eq.~(29) --- here remark \ref{rem:Notationpm}. Their conclusion arises from a misleading notation. From Eqs.~(24) to (36) in Ref.~\cite{Batic2018}, the authors' notation $\bullet_{\pm}$ relates to the asymptotic behaviours as in~\eqref{eq:ExpGrowDec}. Then, the same notation is used in their Eqs.~(37)-(39) to distinguish the pair of complex conjugate values~\eqref{eq:NewQNM}. At this point, Eq.~(40) in~\cite{Batic2018} is misleading. Contradicting their own generic statement after their Eq.~(29), their Eq.~(40) seems to directly attach the $\pm$ sign of the exponential growth/decay to each $\pm$ sign related to imaginary part of $s^{(\pm)}_n$. One concludes --- cf.~ Eqs.~(37), (40) and (43) in Ref.~\cite{Batic2018} --- that picking-up the minus sign in \eqref{eq:NewQNM} leads to the negative sign in Eq.~\eqref{eq:ExpGrowDec} and, therefore, an asymptotic exponential decay. This reasoning appears in their Eqs.~(44) and (52) as well. The authors choose one of the two independent asymptotic behaviour according to the sign of $\Im(s^{(\pm)}_n)$. 

It is already suspicious that $s^{(+)}_n$ is not a QNM in Ref.~\cite{Batic2018}, since $f(\bar{t},x)$ is a real valued function. Despite the inconsistency, we proceed with the authors' arguments and consider the sequence $\{H^{(-)}_{k,-}\}_{k=0}^{\infty}$ obtained from choosing the solution to~\eqref{eq:RecRel} with the decaying asymptotic behaviour at the values $s^{(-)}_n$. Specifically, a generalisation of the Gauss criterion is discussed. Their conclusion is that $\sum_{k=0}^\infty H^{(-)}_{k,-}$ converges when a given inequality is satisfied. This inequality restricts the parameters $n$ and $\ell$ for which $s^{(-)}_n$ could lead to a QNM. The arguments have addressed only condition (II) and there is no guarantee that Eq.~\eqref{eq:RecRel_BC} is satisfied by $\{H^{(-)}_{k,-}\}_{k=0}^{\infty}$ as required by (III). 

To address condition (III), the authors point out around their Eqs.~(53)-(55) that one can straightforwardly iterate the initial seeds~\eqref{eq:RecRel_BC} and the recurrence \eqref{eq:RecRel} forward at $s^{(-)}_n$. Their argument is exactly the statement of Proposition~\ref{rem:ForwardIteration} , i.e., at the values $s^{(-)}_n$ given by~\eqref{eq:NewQNM}, one can construct a solution $\{H^{(-)}_{k}\}_{k=0}^{\infty}$ to~\eqref{eq:RecRel} which satisfies the initial seeds~\eqref{eq:RecRel_BC}. The authors however, do not realise that the sequence $\{H^{(-)}_{k}\}_{k=0}^{\infty}$ obtained by the forward iteration does not lead to the decaying asymptotic behaviour, discussed while arguing condition II.
 
The proof that the sequence fulfilling condition (III) behave asymptotically as required by condition (II) is lacking --- see Remarks~\ref{rem:ForwardIteration_Asympt}-\ref{rem:BackIteration_InitialCondition}, i.e., the authors have not proven that $\{H^{(-)}_{k,-}\}_{k=0}^{\infty}$ and $\{H^{(-)}_{k}\}_{k=0}^{\infty}$ are linearly dependent at $s^{(-)}_n$.

\begin{figure}[t!]
\begin{center}
\includegraphics[width=6.cm]{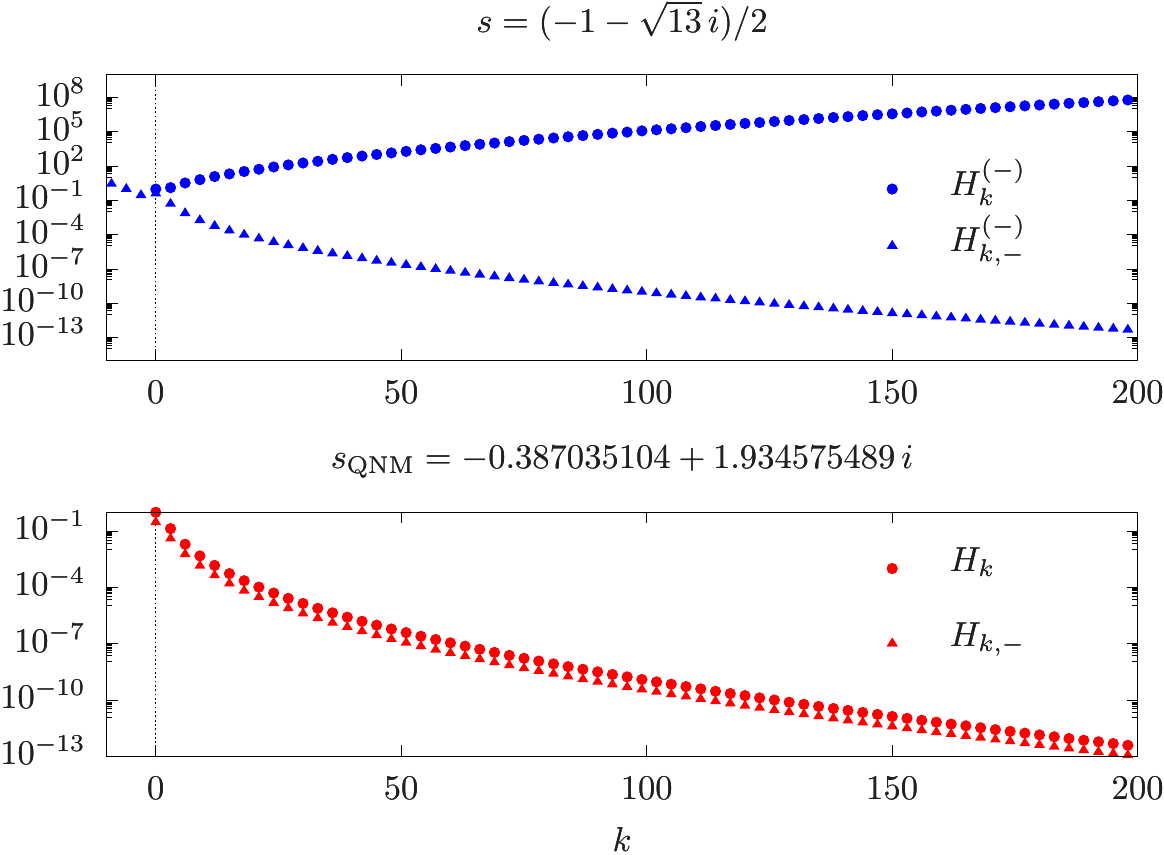}
\includegraphics[width=6.cm]{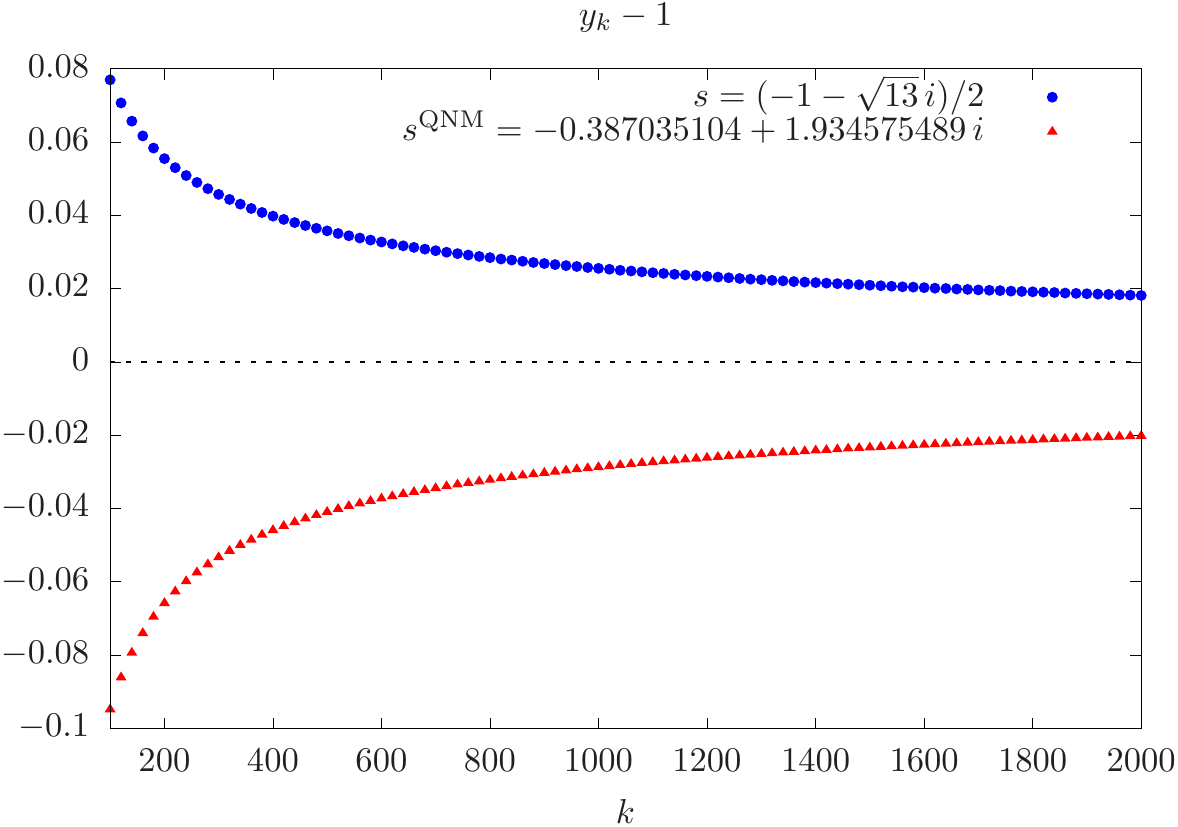}
\end{center}
\caption{
Top Panel: for $s^{(-)}_0=(-1-\sqrt{13}\,i)/2$ (top plot), $\{H^{(-)}_k\}$ grows as $k^\zeta  e^{\kappa\sqrt{k}}$ for $k\gg1$, while $\{H^{(-)}_{k,-}\}$ does not satisfy $H^{(-)}_{k,-}=0$ for $k<0$. For $s=s^{\rm QNM}$ (bottom plot) both sequences become linearly dependent. 
Bottom panel: For $s^{(-)}_0=(-1-\sqrt{13}\,i)/2$ (blue circle), $y_k-1>0$ contradicting the choice made in Eq.~(44) of Ref.~\cite{Batic2018}. For $s=s^{\rm QNM}$ (red triangle) one obtains $y_k-1<0$ as expected.}
\label{fig:FuncQNM_Cinfty}
\end{figure}

\medskip
{\it Counterexamples.}
We verify that $\{H^{(\pm)}_k\}_{k=0}^{\infty}$ constructed for the claimed new branch {\em does not} lead to the decaying asymptotic behavior $y_{k,-}$. Let us fix $n=0$ and $\ell=2$ for $s^{(-)}$ in~\eqref{eq:NewQNM} and follow {\em exactly} page 7 in Ref.~\cite{Batic2018}. Indeed, ``it is straightforward to verify that we can recursively obtain all the unknown coefficients" $H^{(-)}_k$~\cite{Batic2018}. This procedure is exact (no numerical error). One observes that the sequence $\{H^{(-)}_k\}_{k=0}^{N_{\rm max}}$ grows exponentially (blue circles in the top panel of Fig.~\ref{fig:FuncQNM_Cinfty}) according to the theoretical asymptotic $y_{k,+}$ (bottom panel of Fig.~\ref{fig:FuncQNM_Cinfty}). The behavior $y_k-1>0$ contradicts the arguments after (44) in Ref.~\cite{Batic2018} where the negative sign was assumed. If condition (III) is satisfied, then (II) is not.

Alternatively, we construct~\cite{Ansorg2016,Macedo2018} a decaying solution $\{H^{(-)}_{k,-}\}$. We satisfy --- by construction --- the choice for the negative sign in Eq.~(44) of Ref.~\cite{Batic2018}. The initial condition \eqref{eq:RecRel_BC} is not satisfied because $H^{(-)}_{-1,-} \neq 0$ (blue triangles in the top panel of Fig.~\ref{fig:FuncQNM_Cinfty}). If the decaying asymptotic properties required by condition (II) is satisfied, then condition (III) is not met.

Such results are explicit counterexamples to the key arguments of their proof. One identifies the same incompatibility between conditions (II) and (III) for other values of $n$ and $\ell$.

For the sake of comparison, the middle panel of Fig.~\ref{fig:FuncQNM_Cinfty} shows the equivalent results for the first well-known QNM with $\ell=2$: $s^{\rm QNM}= - 0.3870351 +  1.93457545\, i$. The exponential decay for $\{H_k\}$ associated to the minimal solution of the recurrence relation is evident. It is also clear that $\{H_{k,-}\}$ satisfies the boundary condition~\eqref{eq:RecRel_BC} since $H_{-1,-}=0$. In other words, $\{H_k\}$ and $\{H_{k,-}\}$ become linearly dependent at the QNM, i.e., the same sequence satisfies conditions (II) and (III). The results for the QNMs are also displayed in the bottom panel of Fig.~\ref{fig:FuncQNM_Cinfty}. Indeed, for a QNM, one has $y_k - 1 < 0$. 

\medskip
{\it New QNM branches.}
One can extend the arguments in Ref.~\cite{Batic2018} to any value $s^{(\pm)}$ parametrised by~\eqref{eq:QNMCandidate}. Following Ref.~\cite{Batic2018}, one starts with condition (II) and consider Eq.~\eqref{eq:Asymt_y} in its generic form. One argues that ``the case with the plus sign can be disregarded..."~\cite{Batic2018} and we make a {\em choice} to work with the asymptotically decaying sequence $\{H^{(\pm)}_{k,-}\}_{k=0}^{\infty}$. Then, one considers the modified Gauss criterion for the generic $y_{k,-}$ exactly in the same way as in Ref.~\cite{Batic2018}. The convergence of $\sum_{k=0}^\infty H^{(\pm)}_{k,-}$ should be guaranteed when the inequality $2\rho\cos\phi>1$ is satisfied. Finally, one considers condition (III) and concludes: it is ``straightforward to verify that we can recursively obtain all the unknown coefficients", i.e, to obtain the sequence $\{H^{(\pm)}_k\}_{k=0}^{\infty}$. For example, the values $s^{(\pm)}(x) = x\left(-1 \pm\sqrt{3}\, i \right)/2$ with $x\in {\mathbb R},$ $x>1$ meet all the requirements (including the inequality). They are clearly not another new branch of QNMs because the sequences satisfying conditions (II) and (III) are linearly independent.

\begin{figure}[t!]
\begin{center}
\includegraphics[width=6.cm]{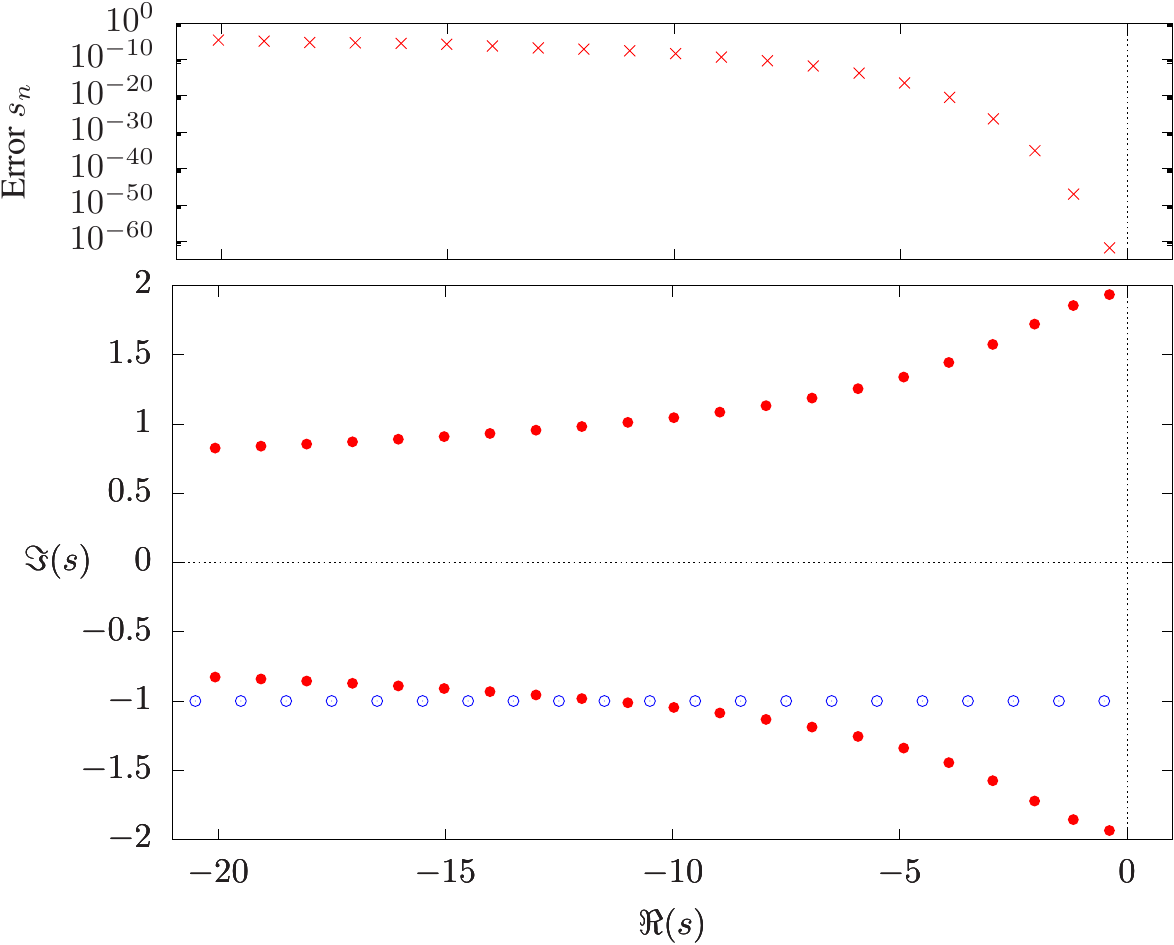}
\includegraphics[width=6.cm]{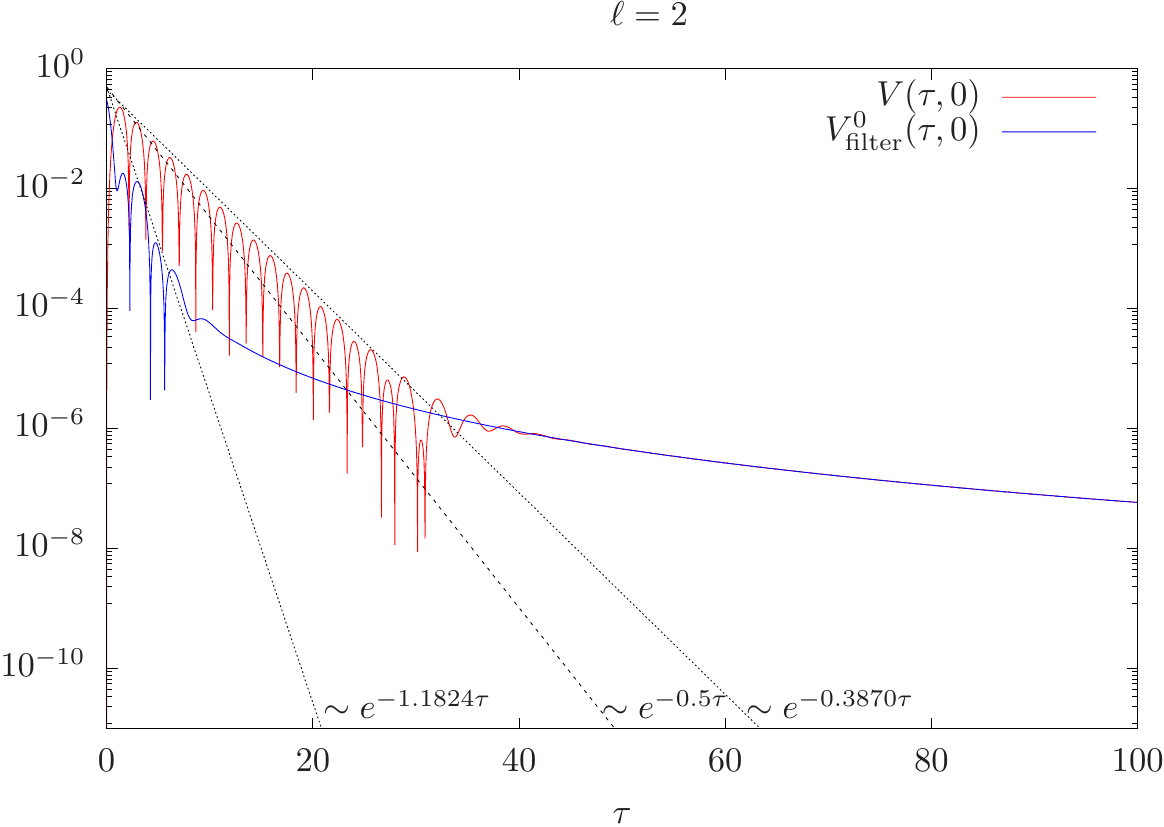}
\end{center}
\caption{Top panel: QNMs as the spectra of the operator ${\mathbf M}$ (solid red circles). The extra panel display the numerical error to each QNM. The values $s^{(-)}_n$ in Eq.~\eqref{eq:NewQNM} (empty blue circles) are not found as eigenvalues of ${\mathbf M}$. Bottom: Time evolution (red) according to~\cite{Macedo:2014bfa}. The filtered field (blue) shows the second QNM $s_1^{\rm QNM}$. No mode with $s_0^{\rm new} = (-1 - \sqrt{13} i)/2$ is detected in the evolution.
}
\label{fig:QNMSpectra}
\end{figure}

\medskip
{\it Alternative methods.} First, we consider the notion of resonances in scattering theory~\cite{DiaZwo17,zworski2017mathematical}. We rewrite Eq.~\eqref{eq:WaveEqHyp} as $- V_{,\tau\tau}  + {\mathbf L}_1 V_{,\tau} +  {\mathbf L}_2 V = 0,$
with ${\mathbf L}_1$ (${\mathbf L}_2$) differential operators of first (second) order, acting only on $\sigma$. After a first order reduction in time, the QNMs are defined as the eigenvalues of the operator
$
{\mathbf M} =\left( \begin{array}{cc}
 0 & {\mathbbm 1} \\ 
{\mathbf L}_2 & {\mathbf L}_1
\end{array}
\right).
$
Figure~\ref{fig:QNMSpectra} shows in red solid circles the eigenvalues of ${\mathbf M}$ for $\ell=2$ with the respective numerical errors. The values of the claimed new branch $s^{(-)}_n$ in Eq.~\eqref{eq:NewQNM} (empty blue circles) lie above the numerical error. They {\em were not found} in the spectrum of the operator ${\mathbf M}$. 

Finally, we consider the direct time integration of Eq.~\eqref{eq:WaveEqHyp}~\cite{Macedo:2014bfa}. Though $V(\tau,\sigma)$ contains information about all the QNMs of the system, the one with slowest damping scale dominates. If \eqref{eq:NewQNM} were a new branch of QNM, the first value would be $s_0^{(-)} = (-1 - \sqrt{13} \, i)/2$. Since $|\Re(s_0^{\rm QNM})|<|\Re(s_0^{(-)})|<|\Re(s_1^{\rm QNM})|$, its contribution to $V(\tau,\sigma)$ would decay faster than the one from $s_0^{\rm QNM}$, but still slower than $s_1^{\rm QNM}$. One attempt to access $s_0^{(-)}$ is to filter the contribution from $s_0^{\rm QNM}$. Reference \cite{Ansorg2016} allows us to independently calculate the amplitude associated to a given $s^{\rm QNM}_n$ and filter its contribution from $V(\tau, \sigma)$. The contribution of $s_0^{(-)}$ within the original signal $V(\tau, \sigma)$ after the filtering is not found.

\medskip
{\it Conclusion. }
We addressed conditions (I)-(III) characterising QNMs in Ref.~\cite{Batic2018}. Using Propositions~\ref{rem:BackIteration} and \ref{rem:ForwardIteration}, respectively, one can always construct sequences $\{H^{(-)}_{k,-}\}_{k=0}^\infty$ and $\{H_k\}_{k=0}^\infty$ that meet conditions (II) and (III). The proof that such sequences are linearly dependent at the values $s^{(-)}_n$ in Eq.~\eqref{eq:NewQNM} is lacking in Ref.~\cite{Batic2018}. In particular, explicit counterexamples illustrate that $\{H^{(-)}_{k,-}\}_{k=0}^\infty$ and $\{H^{-}_k\}_{k=0}^\infty$ are not the same at the values $s^{(-)}_n$ in~\eqref{eq:NewQNM}. Moreover, since condition (I) is valid for any $s$ with $\Re(s)<0$, there should exist further new QNMs if the arguments in Ref.~\cite{Batic2018} were flawless. The frequencies of the new branch were not found in the direct time evolution of the
original wave equation+boundary conditions, and the new values are not eigenvalues 
of the operator associated to the problem. Thus, the announced new frequencies {\em should not} be regarded as quasinormal modes.

\begin{acknowledgements}
I thank Jos\'e Luis Jaramillo for valuable discussions on the topic and Vitor Cardoso
for encouraging the publication of this comment. This work was supported by the European Research Council Grant No. ERC-2014- StG 639022-NewNGR ``New frontiers in numerical general relativity"
\end{acknowledgements}

\bibliographystyle{apsrev4-1-noeprint.bst}
\bibliography{bibitems}

\end{document}